\documentclass{sig-alternate-05-2015}

\usepackage{xspace}
\usepackage[utf8]{inputenc}
\usepackage[T1]{fontenc}
\usepackage{microtype}
\usepackage{threeparttable}

\usepackage[table]{xcolor} 
\usepackage{subfig}
\usepackage{url}
\usepackage{paralist}
\usepackage{siunitx}
\usepackage{booktabs}
\usepackage{tabularx}
\newcolumntype{R}{>{\raggedleft\arraybackslash}X}

\usepackage{tikz}
\usepackage{tikz-qtree}

\usepackage{dblfloatfix} 

\usepackage{todonotes}
\presetkeys%
    {todonotes}%
    {inline,={}}{}%

\CopyrightYear{2016} 
\setcopyright{acmcopyright}
\conferenceinfo{CESI '16,}{May 17 2016, Austin, TX, USA}
\isbn{978-1-4503-4154-7/16/05}\acmPrice{\$15.00}
\doi{http://dx.doi.org/10.1145/2896839.2896844}

\begin{document}
\title{Case Studies in Industry: What We Have Learnt}

\numberofauthors{2}
\author{
\alignauthor
Daniel M\'endez Fern\'andez\\
       \affaddr{Technical University of Munich, Germany}\\
       ORCID ID: 0000-0003-0619-6027\\ 
       \email{daniel.mendez@tum.de}
\alignauthor
Stefan Wagner\\
       \affaddr{University of Stuttgart, Germany}\\
       ORCID ID: 0000-0002-5256-8429\\ 
       \email{stefan.wagner@iste.uni-stuttgart.de}       
}

\maketitle

\begin{abstract}
Case study research has become an important research methodology for exploring phenomena in their natural contexts. Case studies have earned a distinct role in the empirical analysis of software engineering phenomena which are difficult to capture in isolation. Such phenomena often appear in the context of methods and development processes for which it is difficult to run large, controlled experiments as they usually have to reduce the scale in several respects and, hence, are detached from the reality of industrial software development. The other side of the medal is that the realistic socio-economic environments where we conduct case studies -- with real-life cases and realistic conditions --  also pose a plethora of practical challenges to planning and conducting case studies. In this experience report, we discuss such practical challenges and the lessons we learnt in conducting case studies in industry. Our goal is to help especially inexperienced researchers facing their first case studies in industry by increasing their awareness for typical obstacles they might face and practical ways to deal with those obstacles.

\end{abstract}




\section{Introduction}


Empirical software engineering has gained much attention in recent years. The contributions in the community have enabled a shift from a more design-science-driven engineering, where we applied scientific methods to isolated practical problems, to a more epistemology-driven and insight-oriented science~\cite{Basili:2013:Perspectives}. The advances in empirical software engineering are continuously supporting us in the establishment of a reliable software engineering body of knowledge, thus, supporting the scientific progress in our field~\cite{2015-icse-ffc}. However, in contrast to other scientific disciplines governed by precise laws which are often expressible via mathematical forms, such as it is the case in physics, our discipline is to a large extent characterised by exploring complex real-life situations. Those situations involve human beings, each having its own experiences, expectations, fears and beliefs, and their interactions with technology~\cite{runeson09}. In consequence, the observations we make in experimental settings are only valid within those settings. That is, the validity of the theories we build is strongly dependent on the contexts which are difficult to define because of the complex human, economical, technological, and cultural factors involved~\cite{2015-icse-ffc}. Those who neglect the importance of contexts in their research essentially reduce our discipline to its computational layer only.

The importance of exploring software engineering phenomena in their natural contexts gives rise to case study research. In its essence, a case study forms an ``empirical inquiry that investigates contemporary phenomena within its real-life context, especially when the boundaries between phenomenon and context are not clearly evident.''~\cite{Yin03} A case study constitutes therefore a proper way to study an object for which the effects of a treatment are difficult to capture in isolation; for instance, the sensitivity of effort estimation techniques to software development environments following a specific process model. 

However, and as elaborated well by Runeson and H\"ost~\cite{runeson09}, barely any empirical concept is dominated by such a fuzzy understanding of the terminology used and the methods applied, as case study research. This missing awareness for case study research might also cause researchers to believe that case studies are something inferior to experiments. This is, of course, not true and we postulate that experiments and case studies, if both conducted properly, complement each other in scaling up to practice by increasing the degree of realism in the objects under investigation~\cite{Wieringa13}. Both experiments and case studies have their strengths and limitations while case study research is still subject to misconceptions. In response to these misconceptions, Runeson and H\"ost contributed a guideline for conducting and reporting case study research in software engineering, and they provided a valuable framework to foster a standardised understanding on what a case study is and, not less important, how to structure and report it~\cite{runeson09, RHRR12}. They also provide a blueprint to structure the conceptual phases of case studies, including: (1) the design of a study and the instruments in response to specific objectives and goals, (2) the preparation for the data collection, (3) the actual data collection and (4) analysis, and (5) the reporting of a case study. What makes the practical approach to case studies often so difficult, remains that:
\begin{compactenum}
\item Case studies can have different purposes, ranging from exploratory ones where we explore phenomena without predefined expectations or hypotheses to improving purposes where we try to investigate an improvement in a situation. 
\item Case study research is often built upon a combination of different empirical methods. For instance, technical action research~\cite{MA12} and interview research used as a means for the data collection, and Grounded Theory~\cite{AHK11} for the data analysis (e.g. of observation protocols and interview transcripts).
\item The data obtained via case study research can be of both quantitative as well as qualitative nature.
\end{compactenum} 

Despite the conceptual understanding, however, the actual challenges in applying case study research arise in our experience from the ultimate environment of a case study: industrial reality. In case study research, we try to learn from realistic socio-economic environments with realistic conditions including human beings and real-life cases (including realistic models, complexities, and decisions). While the realistic environments constitute the major strength of a case study, they pose a plethora of practical challenges. 

In this experience report, we describe such practical challenges and the lessons we learnt in conducting case studies in industry. Our experience emerges from over 30 case studies we conducted in German industry 
including small ones and ones with up to 30 participants and up to three years of duration, published and unpublished ones, successful and unsuccessful ones. The latter were of special interest when reflecting on our experiences to write this report. Due to the sensitivity of our content and also because we take a post mortem view on our projects, we do not link our experiences to particular projects. Our goal is to share our experiences to help especially inexperienced researchers who face their first case studies in industry by increasing their awareness for typical obstacles they might face and practical ways to deal with those obstacles. 

There are several other valuable sources for success factors and guidelines for collaboration between academia and industry. For example, there are general studies on the objectives in such collaborations and how to best align the ones of the universities and industry~\cite{carayol03}. Furthermore, there are experience reports on the collaboration between universities and companies in projects~\cite{junker15}. Finally, there are also further guidelines -- based on experiences -- on industrial case studies~\cite{Verner:2009fp}. 

In our report, we focus on our own experiences and how one can use them in further case studies. In the following, we report on our experiences structuring them in three topics:
\begin{compactenum}
\item Approaching a case study
\item Conducting a case study
\item Cross-cutting challenges
\end{compactenum}

The report at hands is a follow-up work to the tutorial on in-vivo experimentation given at the \emph{International Advanced School of Empirical Software Engineering} in 2014.

\section{Approaching a Case Study}
\label{sec:challApproaching}

\subsubsection*{Boundaries and Possibilities of Case Studies}

As we discussed in the introduction already, case studies have a distinct role in the empirical analysis
of phenomena, especially in software engineering. The phenomena we analyse are often methods and
development processes for which it is difficult to run large, controlled experiments. If we perform experiments,
we usually have to reduce the scale in several respects and, hence, experiments are detached from the reality of
industrial software development. A case study can be a perfect counterpart which often does not provide
large sample sizes and, therefore, statistical analysis is limited. Yet, a case study, especially in an industrial
context, allows researchers to investigate phenomena in a very realistic context in depth. It is important
to understand these possibilities but also the boundaries of case study research in the planning phase when deciding for the appropriate empirical research methods.

\textbf{Lessons Learnt.} We have had the experience that many researchers have a better understanding
of the possibilities of controlled and quasi-experiments than of case studies. Either, they tend to focus on
quantitative analysis and over-interpret the findings from a usually small sample, or they use their own
toy examples and sell them as case studies. In both cases, they miss out on the benefits an industrial
case study can bring. A case study can provide the most interesting insights, first, if researchers closely collaborate
with practitioners applying the phenomenon under study in day-to-day work (or at least closely resembling
day-to-day work). Hence, ideally, the practitioners only have the researchers as coaches but use the
phenomenon on their own. There, the researchers are the observers only. Second, researchers plan with the usage
of qualitative analysis methods to capture as much of the context, the phenomenon as used, and the outcomes
of its usage. This gives a rich set of data usually impossible to get with experiments.

\subsubsection*{Finding Contacts}

We decided for an industrial case study. We want to conduct it with practitioners as main subjects.
Therefore, we need to find contacts to industrial partners willing to perform these studies. A good
industrial case study includes a non-negligible contribution from the industrial partner, mainly in
the form of time and effort of employees. Finding the right contacts is usually not easy.

\textbf{Lessons Learnt.} We have worked with several strategies that we run in parallel to 
continuously look for new contacts and to reactivate old contacts. First, there exist research
grant programmes designed for a collaboration of universities, research institutes, and industry.
For example, most of the EU Horizon 2020 require such consortia. But also, for example, the German
Ministry for Education and Research funds such consortium projects. Similar programmes exist
all over the world. A straightforward way to come to case study partners is to join such projects and
integrate the studies there --- at best in the proposal already. Second, make use of the companies in
your closer or wider area. While the market for software engineers is large and companies struggle
to hire enough new engineers, the are eager to collaborate. It often starts with a thesis supervised
together before going into bigger case studies. Third, visit practitioner conferences and share your
new methods, insights, and lessons learnt. It is a good way to make new contacts already focused
on a particular topic. You can also create such events for the local industry yourself.
Fourth, stay in contact with your alumni. They might have already a good idea of
what you are doing and can then quickly initiate very focused contacts.

\subsubsection*{Convincing Contacts to Participate}

Finding good contacts to industry is only the first step. The second is to convince them to
participate in a case study. As we have discussed, it requires a substantial contribution from
their side. In general, it is something that goes on top of their normal work. Hence, they need
to be convinced of how the case study is beneficial for them and their work.

\textbf{Lessons Learnt.} Again, probably the easiest way is to run the case study in a
(e.g. publicly) funded project where they also receive funding. This creates freedom for the
practitioners to spend time for it. But also in such projects, it takes convincing as external funding does not compensate for a potentially missing motivation. A practitioner asked to participate in an empirical study to evaluate an effort estimation method developed in context of a PhD thesis might not be as motivated to invest time and effort in that case study as a practitioner seeing potential benefits for her daily work and being asked to participate in a technology transfer project. Technology transfer, same as case study research which is a vehicle to transfer, is a two-way street in which both parties benefit and this needs to be clear to potential participants, but also to the researchers. We need not only to increase (self-)awareness for the role of a case study, but also to apply all good practices of technology transfer to reduce potential barriers impeding the motivation of the participants (see, e.g.,~\cite{DVM15}). In particular, we need to learn to speak the practitioners' language so that there is no communication barrier. Furthermore, we found that it is important to respond to their individual needs in their particular environments. For example, we need to be able to show them that a new code analysis technique, we want to investigate, has the potential to help them find defects faster. Finally, we also learnt that providing feedback early
on and giving them quick results (such as in an iterative approach), helps to keep the participants
interested and improves the results~\cite{2015-icse-ffc}. This also means that we have to be
opportunistic in the subject and the case selection. While it is helpful to be systematic,
researchers need to be prepared to pragmatic changes and to throw parts of their planned selection overboard to get participants on board.

\subsubsection*{Planning a Case Study}

The length of case studies varies, but we made the experience that they often need several
months to get interesting data. Therefore, planning is key to successfully manage the execution
of the study.

\textbf{Lessons Learnt.} We found that there are several aspects that need to be carefully planned at the
beginning of a case study. The researchers need to schedule the necessary meetings and workshops as the
participants might be busy or on holidays otherwise. The necessary resources in terms of people
but also software, hardware, and rooms need to be allocated by the researchers and the participants.
Researchers and participants should jointly define deliverables so that it is clear who has to finish
what and when and who is in charge on quality assurance (and how exactly it shall take place). There should be clear responsibilities for the different parts of the study. Finally, often an non-disclosure agreement has to be made so that it is clear what can be published. Overall, this means that a case study has to be planned in detail with a time and resource plan as well as
quality assurance on protocols, instruments and reports.

\subsubsection*{Dealing with Uncertainty}

An industrial case study has a lot of uncertainties before it starts. For example, it is often unclear
what the characteristics of the stakeholder but also the data  might be. What is the stakeholders' availability, skills, motivation, and
commitment? What will be the quality and quantity of the data?

\textbf{Lessons Learnt.} The best way to address these questions is to structure the case
study iteratively. In particular, take early samples and test them. We always test the used instruments
for data collection and the resulting data quality via pilots. These pilots can first be run internally
with colleagues or students and then on a small number of participants. Furthermore, we recommend
to get to know the subjects as early as possible to reduce the uncertainties and build trust. In the end,
it also helpful to stay flexible and to find ways to use the data you get.

\section{Conducting a Case Study}
\label{sec:challConducting}

\subsubsection*{Properly Characterising the Context}

Properly characterising the context of a case study forms an important step in conducting and reporting a case study, in a way that precisely captures the conditions under which the observations and conclusions hold. The context of a case study supports the transparency and, thus, the reproducibility and the replicability of a case study. Yet, a good context description is difficult to provide and how to elaborate it is, in fact, still subject to ongoing discussions within the community. The reasons why it is so difficult to provide a rich context description include that there is no clear understanding which factors might be important, how these factors can be measured, but also which factors can be reported within the limits of non-disclosure agreements. 

\textbf{Lessons Learnt.} A context description should provide a holistic picture of the case study going beyond case and subject descriptions for which one wants to draw conclusions, e.g.\ by providing an overview of the software development project and the company in which the case study was conducted. We found it useful, as a first step, to take existing case studies with similar objectives as orientation while asking ourselves: ``What information would I myself need to understand and replicate the study under the assumption it would yield the same or similar results?'' One reference we found useful to take into consideration stems from the field of software process models and their tailoring. Such tailoring models include organisational, human, and technological factors that play an important role when customising the way of working. Hence, they serve as an orientation to characterise the industrial environment as a whole for a case study, too. An exemplary model can be taken from~\cite{kk2013a}.

\subsubsection*{Communication}
Communication plays a vital role in case study research, especially if conducted in large companies with complex organisational structures and strict communication lines to follow. Possible problems we experienced to occur and, thus, impede the success of a study include, if not limited to: No clear communication and reporting lines have been defined, the case study itself has not been properly communicated within the organisation, or it might have even been delegated to someone not interested at all in the study. 

\textbf{Lessons Learnt.} We found it crucial to define clear lines for communication, reporting, and also for escalations right from the beginning when planning a case study. One aspect we consider very important, however, concerns the stage when executing the case study: Take no shortcuts! Playing by the rules of the organisation is not only a matter of loyalty, but it is the foundation for building trustful environments. Besides the organisational aspects, it is important to ensure transparency by clearly communicating the purpose and the goals of a case study itself and also which results will be reported (and how they will be reported to whom). Finally, one thing often underestimated is to consider the fine but important difference between a sponsor of a case study and a champion. The latter is often hard to identify, but working with someone with an intrinsic motivation at project level and reflecting her own needs and interests in the study opens many doors which otherwise might remain closed.

\subsubsection*{Defining the Proper Instruments}
The instruments used for the data collection can range from tool support to extract and analyse code comments to questionnaires used in interviews or surveys. Realising too late that the instrumentation is not sufficient to satisfy the research objectives can bring the whole study to a dead end. One problem is that we might not have thought through our research questions. The far more complicated problem is, however, that we often have to make assumptions about the population in advance which can turn out to be wrong w.r.t.\ size and quality (i.e. we have to deal with uncertainty). Sometimes, we even face a luxury problem where the data we collect turns out to be richer than expected as it can answer questions we never had in mind. For instance, in one study~\cite{MWLBC10} we originally wanted to analyse different patterns in requirements engineering by analysing the artefacts created. While executing the study, we got access to data that allowed us to also track change requests and the effort spent, thus, giving us the possibility to evaluate the efficiency of the patterns to a certain extent. There, we decided to extend our case study with additional research questions during its execution. Some changes, as this one, have no effect on the validity as we can build our analysis on top of the previous one; other changes, however, might lead to having to stop the study if they imply modifying the instruments already used.

\textbf{Lessons Learnt.} We experienced two things to be important. First, we need to approach the design of a study and its instruments as we would approach the design of a software product: iteratively and preferably in small steps. We believe that there is no such thing as a purely idealistic top-down approach (from goals over research questions to measurements), nor is there a seemingly pragmatic bottom-up approach (``show me the data and I'll show you the questions''). The ideal approach is somewhere in the middle. Ideally, we should start asking small questions we are realistically able to answer; and we then should increase our study design step-by-step until reaching our general objective while refining our goals along the design of the instruments. Second, to reduce the risks arising from uncertainty, we should test our design and the data quality, for example, by running a pilot of a survey. This allows us to test not only the design (e.g.\ whether the questions in a questionnaire are understandable by practitioners), but also the data analysis techniques which we planned to use by applying them on the test data. One way we found also useful to calibrate the instruments is to conduct a small experiment with students first. 

\subsubsection*{The Data Collection Might Take Long}

The data collection in a case study might take long and sometimes longer than expected. One reason is that we might depend on practitioners  who have to schedule their involvement besides their daily work. Another reason might simply be to run the data collection as part of a longitudinal study that might inherently span several years. 

\textbf{Lessons Learnt.} We believe that there is no universal solution to this challenge, but there are three lessons we personally learnt to be important: We need to be patient, we need to plan buffers in our time schedules, and we should never push our respondents. 

\subsubsection*{Working with Sensitive Data}

Working with real life data that emerges from industrial environments means working with highly sensitive data. Exposing this data to the public without clearance is therefore not only problematic from an ethical perspective, but also from a legal one. On the other hand, reporting on a case study without reporting the data violates the principles of scientific working as it is asking the scientific community for too much credit. We therefore need a balance between the researchers' duty to publish the results while preserving the companies commercial interests~\cite{AP01}, i.e.\ we need to find the proper way to ensure transparency of a study while staying within the limits of non-disclosure agreements. 

\textbf{Lessons Learnt.} The most important thing we experienced is to clarify the rules in advance including what you need for the study, which data can be stored and where it can be stored, what can be reported internally within the company and externally to the public, and finally how the data can be reported. The latter implies to clarify how much abstraction from information is necessary for the company while still useful for a reproducible reporting. Companies probably do not want to see a publicly accessible report that shows from which projects and, thus, customers the data exactly emerges while the information actually important to our analysis might not be per-se problematic. They probably also do not want a report to only show what does not work in their projects, but to make explicit that the analysis has been conducted in an endeavour to improve the situation in the company.  A proper reporting of the study and its data along the rules of the company also means to plan for review cycles and acceptance phases when reporting on case studies (internally as well as externally). Finally, it is also important to include industry partners as co-authors where possible and where reasonable. They can take an active role in the reporting process, especially within the organisation, and, even more important, it gives credit to their contribution when deserved (respecting the Vancouver Protocol).


\subsubsection*{Skills Matter}

The quality of the outcome of a case study strongly depends on the skills of the involved researchers as these affect the overall lifecycle from the design of the study over the data collection to the analysis and reporting. We experienced especially the data collection to include a variety of challenges that might possibly distort the results, including, for instance, biased researchers trying to sell their own PhD topic, weak moderation and listening skills, or hidden agendas of the respondents potentially resulting from a lack of trust. For instance, in a study we conducted as part of a technology transfer project, we interviewed stakeholders to better understand their needs in context of enterprise architecture solutions. There, one PhD student, probably without realising, continuously suggested answers that would be in tune with her own expectations. Needless to mention that such a bias is easy to detect and criticise from the sidelines when observing as a moderater of the interviews; but it might be difficult to detect and prevent when conducting the interviews ourselves.

\textbf{Lessons Learnt.} Skills are an asset that researchers gain through practice and experience. In our experience, one important aspect for the improvement of the skills is the (self-)awareness of a potential lack of skills. We believe it is especially important for inexperienced researchers to actively approach experienced colleagues and involve them in the design of the instruments and in the data collection procedures (e.g. by inviting them to conduct or at least moderate the first interviews). A continuous exchange and practicing fosters the learning curve necessary for self-improvement with respect to the skills -- important for an accurate study but also to build trustful environments -- and also to get an unbiased and objective view on our own studies.

\subsubsection*{Analysing Qualitative Data}

Qualitative data plays a vital role in case studies and allows us to tell the story behind statistical figures, i.e.\ knowledge is more than statistical significance~\cite{RHRR12}. Especially when reasoning about phenomena or searching for explanations for phenomena, we found it useful to rely on qualitative data gathered, for instance, via interviews. Such data provides us with practitioners' expert views and their experiences and expectations, but also with very personal beliefs and psychological facets. However, as much as qualitative data captures personal views, its analysis and interpretation is also influenced by the views of the researchers (let alone our own mental models). Qualitative data is inherently subjective and so is its interpretation as it is also to a large extent a creative task~\cite{WM15}.

\textbf{Lessons Learnt.} When analysing qualitative data, we found it important to work in teams, no matter the level of expertise and experience of the researcher (researcher triangulation). This reduces the threats to the internal validity. To increase the reproducibility of qualitative data analyses, we also experienced it to be important to document the rational for every interpretation we make. For instance, when coding textual data, we often add a rational for the chosen codes to the text fragments (see~\cite{WM15} for examples). Further, as the analysis of qualitative data is pervaded by subjectivity, it is important to validate, where possible, the results by consulting the practitioners who provided input for the data, e.g.\ by discussing resulting models with them. Finally, although often neglected, it is imperative to be careful when drawing conclusions from qualitative data, e.g.\ by over-generalising or by interpreting too much into the data. This is also one reason why we believe in the importance of reporting objectively the qualitative data while disclosing it (even if anonymised) to the public encouraging readers to run their own analyses and make their own interpretations.

\subsubsection*{Dealing with Moving Targets}

Moving targets are something natural in case study research let alone because case studies tend to take long while often not keeping pace with changing needs in industry. The problems come, however, not with the moving targets themselves but if not detecting them on time as it leaves limited possibilities to properly react to them. 

\textbf{Lessons Learnt.} One key to properly deal with moving targets is, in its essence, the same as known from software development giving rise to the agile movement: fast feedback cycles~\cite{2015-icse-ffc}, i.e.\ continuously providing and getting feedback in interaction with the industry participants. In our experience, it is important to plan and agree early on key questions to be answered while planning for new questions unknown in advance and potentially arising along a study. This allows to focus on the core parts of a study while preserving the necessary flexibility to meet both the interest of industry participants and the researchers involved.

\section{Cross-cutting Challenges}
\label{sec:challTransversal}

\subsubsection*{Beware the Ivory Tower}

A problem that can arise before or during a case study is that the management supports conducting
the study, but there is no support in the actual development projects. In one case, we were developing a model-based requirements engineering approach for embedded systems in the automotive domain. After three months of working with the responsible process engineers, we realised that we would not get any access to the engineers supposed to apply the approach leaving us in the situation that we had been working on a solution for a context-specific problem we would never be able to completely understand. This can happen when researchers discuss the study only at the management level of the company. They often find it interesting, want good relations to the university, or engage in a previously agreed project phase (e.g.\ an evaluation phase during a technology transfer project). Yet, although agreeing on the project setting, the topic might not be interesting or relevant at project level.

\textbf{Lessons Learnt.} First, researchers can address this problem by writing down a clear study plan
including what subjects and cases are needed. All involved parties then formally agree on this study plan.
Second, a formally agreed on plan does not mean it will work. Researchers should therefore talk to intended subjects
as early as possible to keep them informed and get their feedback. Even better is an active supporter at the
development team level who has an own interest in the study (the champion).

\subsubsection*{Beware the Sample}

Samples in industrial case studies tend to be smaller than, for example, in experiments or in studies relying on publicly available data. The focus of case studies is usually set on studying details of realistic, company-specific models and their effects rather than studying isolated facets in a large amount of models. It is natural that we analyse one or two cases when they each constitute a whole software development project and it is also natural that we work with a low number of subjects when we work, for example, with process engineers. Low sample sizes are in our opinion not a problem in case studies, a low quality of the data, its analysis or reporting, in turn, is. 

\textbf{Lessons Learnt.} The notion of quality is particularly fuzzy in case studies, but two criteria we believe to be important to beware are (1) the level of detail when reporting on cases and subjects, and (2) their representativeness: ``Is this role representative for the context? Are her views representative? Are these extreme views worth treating as an outlier or maybe worth further exploring in detail?'' Both criteria eventually allow us to properly draw conclusions, identify and reason about interesting phenomena, and finally compare our results in relation to existing evidence. We also experienced that many researchers not familiar with the natural context of industrial case studies criticise, for instance, low sample sizes. We believe, however, that it is close to impossible to analyse both realistic models and large sample sizes in great detail.

\subsubsection*{Drawing Proper Conclusions}

Also in response to low sample sizes comes often the difficulty to draw proper and interesting conclusions. This can range from seeing phenomena where there are none over making interpretations with no solid empirical basis to over-selling results. We often experienced especially inexperienced researchers trying to over-sell their results even though their results might already have been interesting. 

\textbf{Lessons Learnt.} We are convinced that every empirical observation is worth reporting. We also believe, however, that it is especially imperative in case studies  to be as accurate, critical, unbiased, and objective as possible. Interpretations should be clearly stated as such and where possible, existing evidence should be taken for explanations of observations rather than own (subjective) interpretations.

\subsubsection*{The Quest for Universal Truth}

The need for robust theories in software engineering research has been subject to discussions in the community for a while~\cite{jacobson2009we}, and we are yet at the beginning of building a reliable body of knowledge for our field relying on such theories. Yet, one pitfall we can often observe especially inexperienced researchers to step into consists of the generalisation of their observations. Simple fact is that there is no such thing as universal truth, i.e. theories that are valid without particular context information, because of two things: (1) A large extent of software engineering phenomena are subject-dependent (subjectivity), and (2) reality cannot be completely captured in a system of confounding variables (see e.g.\ \cite{MMFV14} for a richer discussion). Truth will therefore be always something relative to its context whereas the context is something too often forgotten when reporting the study. 

\textbf{Lessons Learnt.} We believe that a clear and rich description of the context of a case study constitutes a crucial if not the most important step in forming theories out of empirical observations. Aiming for trustworthiness and transparency increases the reliability and reproducibility of observations and, thus, make even studies with small but realistic cases valuable to expanding our body of knowledge. We also believe that one important asset in building robust theories consists of the awareness that this cannot happen in one step, but that it takes the joint effort of a whole community.  Prerequisites for this are (1) the appreciation of replication studies, also and maybe especially ones that cannot confirm existing theories as they support exploring fail conditions, and (2) the direct and indirect support of replications by disclosing the data and by constantly making explicit the relation to existing evidence.

\subsubsection*{(Data) Openness and Transparency}

Openness and transparency has recently become subject to public discourses\footnote{See, for instance, the Nature initiative on reducing the irreproducibility (\url{http://goo.gl/drgtH}) or the Peer Reviewers' Openness Initiative (\url{https://goo.gl/hYCSOt}).}. In our opinion, there exist many reasons potentially impeding researchers from disclosing their data. Some researchers might not want to pass through exhausting clearance processes of a company; some might not be willing to invest the effort necessary for anonymising the data and making it understandable to others not involved in the study; some might not want to disclose the data before their study is eventually published while the intention to disclose it then might starve later on in their To-Do list; others might fear an independent re-analysis of the data or a replication of the study that could yield different results. However, as discussed above, no matter what the reason to keep the data closed might be, openness and transparency are prerequisites for the reproducibility and replicability of the study and for the reliability and trustworthiness of the results. 

\textbf{Lessons Learnt.}  There is no universal solution to this challenge, but it is clear that the scientific progress in our field, same as in other fields, depends on transparency. It should therefore be our natural duty to disclose our data to publicly accessible repositories\footnote{One prominent repository for sharing software engineering research data is the PROMISE repository under~\url{http://openscience.us/repo/}}. This means that the data needs to be anonymised to the extent necessary to obey non-disclosure agreements and it needs to be enriched with information so that external researchers can work with it. Should disclosing the data not be possible, we need to clearly report on the reasons, and the threats to the credibility need to be compensated by reporting on the information necessary to reproduce the findings of the study.

\subsubsection*{Properly Reporting the Results}

As any research, case studies need to be properly reported. We usually focus on writing up the
design and results suitable for a conference or journal. However, especially in industrial case 
studies, also a dissemination into practice should be supported. The results are
hopefully useful to a broader range of software engineers whereby the reporting of a case study should be always done in a way and with the mediums accessibly by the expected audience.

\textbf{Lessons Learnt.} Nowadays, we have a variety of mediums at our disposal allowing us to reach out to a broad audience: from informal presentations and blog posts, over technical reports to peer-reviewed publications. As a first step, the results should be disseminated inside the company where the case study was conducted. This can be achieved by small presentations to the management as well as a broader presentation if there are regular presentation slots at the company. Second, a write-up of the whole report on the case study goals, design
and results together with instruments and data should be made publicly available, e.g., as a
technical report and/or on an open platform. Finally, the report should also be sent
for peer review at a conference or journal to add credibility and visibility. Especially for that, we
advise to adhere to established guidelines~\cite{runeson09}.

\section{Conclusion}
\label{sec:conclusion}

In this experience report, we summarised challenges and lessons we learnt while conducting case studies in industry. Those experiences and views emerge from various case studies conducted in academia-industry collaborations. Reporting those experiences from a researchers perspective shall support other researchers, especially young ones, facing their first case studies by increasing their awareness for typical obstacles they might face. Where possible, we gave practical advice on how to deal with the obstacles. Below, we summarise those success factors for approaching and conducting case studies in industry we believe to be important. 

\begin{center}
\begin{tabular}{|l p{7.3cm}|}
\hline
\rowcolor{gray!50}
\multicolumn{2}{|c|}{\textbf{Success Factors for Case Studies in Industry}}\\
\hline
1. & Carefully plan your case study with sufficient space for uncertainties \\
2. & Design the study iteratively in small steps \\
3. & Test the design and the intended data analysis techniques via pilots  \\
4. & Understand and document the context \\
5. & Make transparent the goals, purpose, context, and intended reporting of the data to the participants \\
6. & Follow the principles of technology transfer to reflect industry needs and play by the organisational rules \\
7. & Continuously provide and get feedback \\
8. & Be flexible and pragmatic to changes \\
9. & Continuously practice your empirical and social skills \\
10. & Disclose your (anonymised) data to the public for the purpose of credibility and replicability \\
11. & Work closely with the industry participants and in teams with experienced researchers \\
12. & Be unbiased, accurate, and critical \\
13. & Be patient and never push \\
14. & Report for the audience and choose the medium accordingly while relying on established guidelines \\
\hline
\end{tabular}
\end{center}

Those success factors for conducting case studies in industry are our very personal ones as they result from our personal experiences reflecting our own views. There are further factors we might not have thought about or we thought where not so important. Other researchers might come up with a different list as they might have made different experiences. Our hope is also to encourage other researchers to share their experiences on conducting case study research in industry but also on other empirical research methodologies to strengthen the level of expertise in our field. This paper shall provide a first step in this endeavour.

\bibliographystyle{abbrv}
\bibliography{Literature}
\end{document}